\documentclass[11pt]{article}
\usepackage{times}
\usepackage{geometry}
\usepackage[utf8]{inputenc}
\usepackage{enumitem,amssymb}
\usepackage{graphicx}
\usepackage{fancyhdr}
\usepackage{aas_macros}
\usepackage{mdframed} 

\usepackage[authoryear]{natbib}
\setcitestyle{authoryear,open={(},close={)}}
\mdfdefinestyle{theoremstyle}{
innertopmargin=\topskip,}
\mdtheorem[style=theoremstyle]{lrptextbox}{}

\begin{document}

\Huge
\centerline{\bf Cosmology in front of the background:}
\smallskip
\LARGE
\centerline{\bf studying the growth of structure at CMB wavelengths}
\bigskip
\normalsize
\centerline{A white paper submitted as part of the 2020 Canadian Astronomy
Long Range Plan process}
\bigskip

\noindent
{\sl Douglas Scott (UBC),
J. Richard Bond (CITA),
Scott Chapman (Dalhousie),
Dagoberto Contreras (York),
Michael Fich (Waterloo),
Mark Halpern (UBC),
Gary Hinshaw (UBC),
Ren{\'e}e Hlozek (UofT),
Jonathan Sievers (McGill)}
\medskip

\section*{Executive summary}
Canada has thriving communities in CMB (cosmic microwave background) studies,
cosmology and submillimetre (submm) astronomy,
with involvement in many facilities that featured prominently
in previous Astronomy Long Range Plans.  The standard cosmological model
continues to be well fit using a small number of parameters.  No one expects
this model to be complete and so we need to continue to challenge it with
data; moreover, it does not explain how galaxies and other structures form.
So, how do we improve
the precision of our understanding of structure formation within this model?

Wavelengths from the microwave to the submm will be particularly
fruitful for answering this question.  That's because, in addition to the CMB
anisotropies, there are other signals that can be extracted from large maps at
these wavelengths -- particularly
the cosmic infrared and submm backgrounds, the thermal and
kinetic Sunyaev-Zeldovich effects, and CMB lensing.  Such signals carry a
wealth of information about the cosmological model, as well as how dust, gas
and star-formation evolve within dark-matter halos.  Cross-correlations between
these signals and those coming from the radio, optical and X-ray surveys,
will provide even more information.

Canadians are already members of teams for several related facilities and
are working to be involved in others.  In order for Canada to be fully engaged
in exploiting the detailed information coming from these cosmological
signatures, it is crucial that we find the resources to participate
competitively in a combination of projects currently being planned.
Examples include CMB-S4, CCAT-prime, AtLAST, a new
camera for JCMT, balloon projects such as BFORE and a future ambitious CMB
satellite.



%


\section{Introduction}

Canadian astronomers have a particular interest and expertise
in wide-field surveys.  Among such surveys, the most ambitious
ones are carried out as major collaborative efforts with
astronomers from other countries.
The widest (and perhaps highest impact) of these surveys have come from the
{\it WMAP\/} satellite at microwave frequencies \citep[e.g.,][]{WMAP9},
and from the {\it Planck\/} satellite covering microwave-to-submm
frequencies \citep[e.g.,][]{Planck2018_I}.  Although the beamsizes
are modest, these data sets cover the entire sky in 14 bands from
20 to $900\,$GHz.

We all know that these maps can be used to extract the primary cosmic
microwave background (CMB) signal,
sometimes called the ``baby picture'' of the Universe, whose statistical
properties are currently the dominant way of constraining the parameters that
describe our Cosmos -- this is discussed in a separate white paper
\citep{cmb_wp}.  To get to the ``background'' we must first extract the
``foregrounds'', which mostly come from the Milky Way Galaxy, from the
inside of which we observe the CMB.  It is also well known that
these foreground signals can teach us about the large-scale properties of
our Galaxy \citep[e.g.,][and references therein]{Planck2018_XII}, another
science area in which Canadians excel.

\begin{figure}[htbp!]
    \centering
    \includegraphics[width=0.9\textwidth]{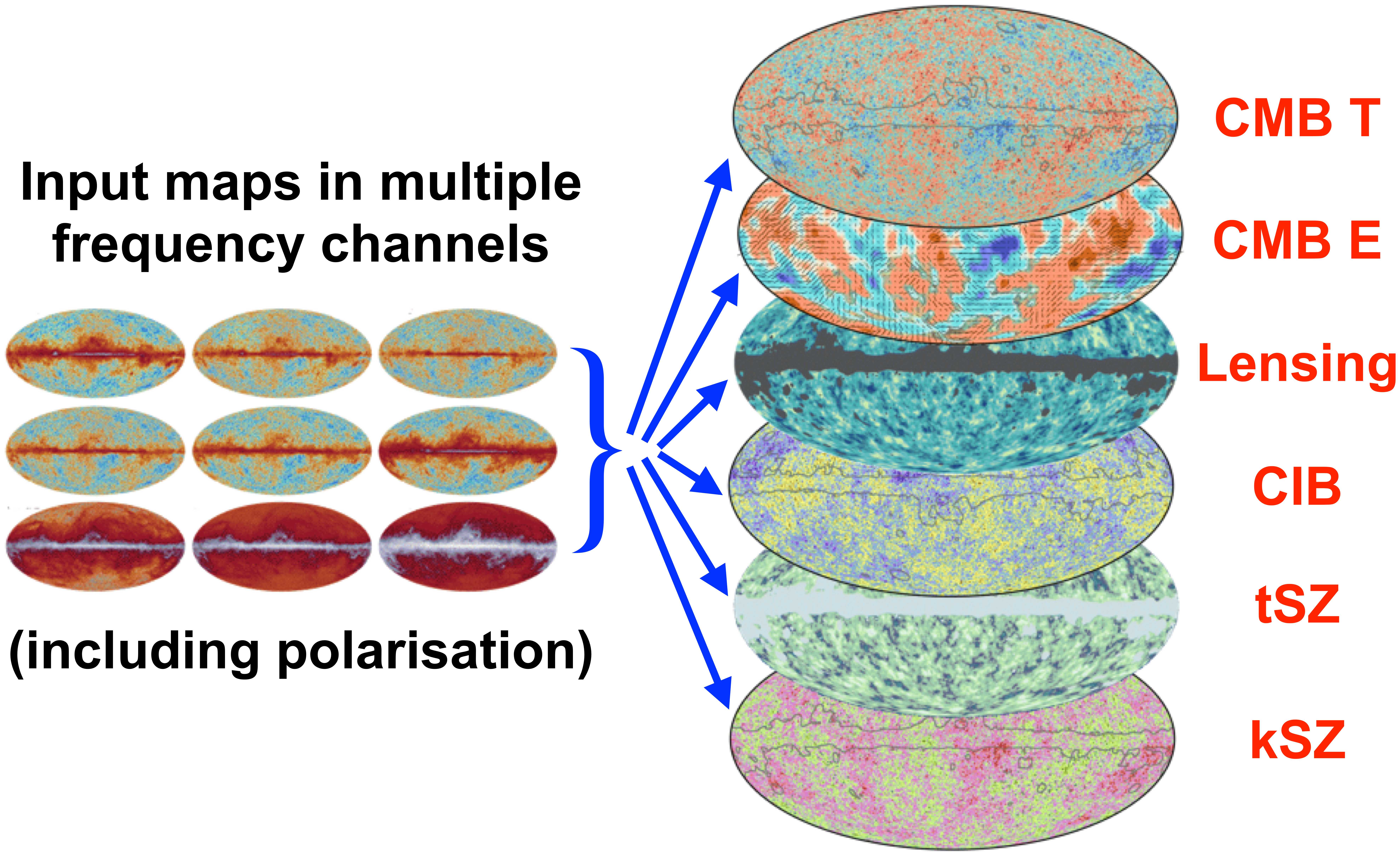}
    \caption{{\sl CMB surveys will map large fractions of the sky in several
    distinct frequency channels (potentially including Stokes $Q$ and $U$
    polarization, as well as $I$), indicated schematically here.
    These can be combined to yield maps of Galactic foregrounds (not shown),
    together with the primary CMB temperature and polarisation $E$-mode
    maps.  In addition it is possible to extract several other signatures
    of structure formation at redshifts much lower than the CMB
    last-scattering epoch, in particular from
    CMB lensing, the cosmic infrared background, and the thermal and
    kinetic Sunyaev-Zeldovich effects.}
\label{fig:maps}}
\end{figure}

What is perhaps less well known is that there are several ``secondary''
signals in these CMB maps, which are of growing interest.  For many CMB
experiments these give ancillary science motivations; however, for some
surveys planned at similar wavelengths such ``secondary'' signals are in
fact the primary science goals.  The CMB anisotropies (in both temperature
and polarisation) were largely formed at the last-scattering epoch for
CMB photons, which was at a redshift $z\simeq1100$,
when the Universe was about 400{,}000 years old.
However, there are several signals that arose at lower redshift, when
structure was forming in earnest.  These additional cosmological signatures
come from gravitational lensing, scattering processes and emissivity in
sources.  With sufficient frequency discrimination, CMB experiments can
produce several distinct maps of these ``non-CMB'' signatures (see
Fig.~\ref{fig:maps}).

Determining the cosmological information content from Gaussian fluctuations is
just a matter of counting modes.
Any map that is constructed up to a resolution corresponding
to multipole $\ell_{\rm max}$ contains ${\sim}\,\ell_{\rm max}^2$ modes.
The usual divergence-type (or $E$-mode) polarisation thus effectively doubles
the number of modes that are probed by CMB experiments that have polarisation
sensitivity (more if we can also measure $B$ modes; \citealt{SCNM}).
The ``non-CMB'' maps are currently noise-dominated (rather than
cosmic-variance-dominated), but that will change with
more sensitive mapping.  Hence the sheer amount of information telling
us about cosmology and structure formation will grow by at least a factor
of a few.  However, not all information has equal value and we can expect
explicit new kinds of information to come from some of these additional modes
that are being probed -- for example in constraints on dark-energy evolution,
dark-matter properties, feedback processes in structure formation, etc.
In addition these secondary signals are {\it not\/} Gaussian, and hence we
can extract more than just the power spectrum.  Indeed by combining with
other probes of the same low-redshift Universe, we can simultaneously
constrain the cosmological model {\it and\/} learn about how the baryonic
components -- stars, star formation, cold gas, hot gas and dust --
occupy dark-matter halos.

\section{CMB experiments}

There are many new and planned projects to map the sky at wavelengths
from the microwave to the submm.  A (non-exhaustive) list of
currently ongoing experiments includes ACTPol, BICEP/Keck, CLASS, SPIDER and
SPTPol.  In terms of future wide-field
surveys that will provide valuable information on CMB ``secondaries'', there
is AdvACT and SPT-3G, with
Simons Observatory \citep{SO} being an ambitious ground-based program that is
currently under construction, and then there is the even more ambitious
CMB-Stage 4 \citep{CMBS4_ScienceCase} project that is in the planning stages.
Large cameras on single-dish mm and submm telescopes
(like IRAM or the LMT) perform surveys that are effectively like CMB
mapping at higher angular resolution -- opportunities here include
a new detector array for JCMT, the CCAT-prime project
\citep{CCATp} and plans for ATLAST \citep{AtLAST}.  There are also ideas
for surveys from balloons (which can go to even higher frequencies),
like BFORE \citep{BFORE}.  And the ultimate concept is a very powerful
future satellite, such as the ``Backlight'' idea \citep{Backlight}.

To pick on one example here, AtLAST is a concept for a 50-m single-dish
submm telescope on the Atacama Plateau, with a continuum camera having
a thousand times the mapping
speed of current facilities.  By adding powerful spectroscopic capabilities
it will be possible to produce a ``submm SDSS'', revolutionising our
understanding of the evolution of star-formation in the Cosmos.

CCAT-prime, being at a higher altitude even than ALMA, can extend surveys
to higher frequencies, with a telescope that is largely dedicated to
wide surveys.  CCAT-p is a 6-m telescope that is currently being built,
for completion in 2023.  Its first-light camera will cover five bands in the
mm-to-submm range, enabling much of the science discussed here.
Canadians are already part of the project, and trying to secure funding to
make significant contributions \citep[see the white paper by][]{single_dish_wp}.

Another dimension in such imaging surveys is the inclusion of polarimetry.
Canadians have already been involved in such instruments, through POL-2
on JCMT, the {\it Planck\/} satellite and BLAST-pol
(the polarisation-sensitive version
of the successful BLAST balloon project).  A new camera envisioned for
JCMT would have polarimetric capability, in addition to dramatically
increased mapping speed.  Many of the current and future CMB experiments
(for example the {\it LiteBIRD\/} satellite) are
focusing on polarisation because of the high-priority search for primordial
curl (or ``$B$'') modes in the polarisation pattern.  This means that
we will have extremely deep intensity maps, with complementary polarisation
information.

Let us turn to an example of such a future CMB experiment.
``{\it CMB-S4 is envisioned to be the ultimate ground-based cosmic microwave
background experiment, crossing critical thresholds in our
understanding of the origin and evolution of the Universe, from the highest
energies at the dawn of time through the growth of structure to the
present day\/}'' \citep{CMBS4_wp}.  This so-called ``Stage~4'' experiment
has as its main science goals a measurement or upper limit on the amplitude
of primordial $B$ modes at the $10^{-3}$ level and constraints on light-relic
particles
that contribute at the level of only a few percent of a standard-model
neutrino.  In order to achieve these goals, CMB-S4 will make an extraordinarily
deep map of about 1000\,${\rm deg}^2$, as well as a wide map of about half of
the sky.  These regions will be covered in several bands between 30 and
300\,GHz (1\,cm to 1\,mm in wavelength), with a large fraction of the detectors
being around 150\,GHz (2\,mm), where the beamsize will be about $1\,$arcmin.
This means that in addition to the inflationary and dark-Universe science
goals, CMB-S4 will make exquisite maps of the CMB secondary signals, thus
tracking how structure forms at relatively low redshifts.

The ultimate project currently being discussed is a large-aperature CMB
telescope in space, with a huge number of detectors, where a large part
of the science goals are exactly what we are describing here -- extracting
the cosmological signals that lie in front of the CMB.  A plan for this
``CMB Backlight'' mission was submitted to ESA's ``Voyage 2050'' process
\citep{Backlight}.

\section{Signals in front of the CMB}

So what are these ``secondary'' signals that come out of CMB surveys, and what
can they be used to tell us about structure formation?

\subsection{Gravitational lensing}

The CMB provides the oldest photons that we can detect.  The paths of these
photons are diverted by the gravitational influence of the matter that they
pass through on their journey towards us.  This induces correlations in the
maps of CMB $T$ and $E$ modes, which can be extracted and used to produce a
map of the large-scale lensing potential
\citep[e.g.,][and references therein]{Planck2018_VIII}.
Polarisation maps will be particularly powerful in future, since, as the
sensitivity continues to be enhanced, they are
lacking most of the foreground signals that can contaminate temperature maps.

The statistics of CMB lensing maps contain information that can be exploited to
constrain dark energy and modified gravity, as well as the sum of the masses
of the neutrino species \citep[e.g.,][]{CMBS4_ScienceBook}.
The lensing map produced by CMB experiments gives
an image of the dark matter along the line of sight to the last-scattering
surface.  The main contributions come from structures at $z\,{\sim}\,1$--3,
which are also probed in wide-field surveys in other wavebands (optical, radio,
etc.).  Since lensing traces the effect of {\it all\/} the matter, then
comparisons with surveys that are sensitive to the baryons (stars, gas, etc.)
can help us understand the relationship between baryonic matter and
dark matter.

As well as weak lensing, there are strong-lensing signals to extract.
The steep slope of the flux distribution of submm galaxies means that it
is an efficient waveband for finding strong lenses.  Such studies
have already begun, using {\it Herschel} \cite{Herschel_lenses},
{\it Planck\/} \citep{GEMS}
and SPT \citep{SPT_lenses}, for example,
but with relatively modest sample sizes.  There should be many thousands in
future surveys.  Investigating the abundance of such lenses can in principle
measure background cosmology, while detailed studies of substructure can
be used to constrain dark-matter properties \citep{Hezaveh}.

\subsection{Thermal Sunyaev-Zeldovich effect}

The thermal Sunyaev-Zeldovich (or tSZ) effect is the upscattering of CMB
photons as they pass through hot, ionised gas, in particular in the halos
of galaxy clusters.  The effect is essentially redshift independent, and hence
can be used to find clusters at higher redshifts than other techniques.
Catalogues of SZ-detected clusters already contain 
${>}\,1{,}000$ objects \citep{Planck2016_XXVII}, new surveys will find
${>}\,10{,}000$, and eventually surpass 100{,}000 (for CMB-S4)
or even 1{,}000{,}000 (for the Backlight).

The counts of clusters are a sensitive probe of background cosmology, and
hence parameters (including dark-energy and neutrino physics) can be
constrained through estimation of $N({>}M,z)$.  Of course this can only be
done effectively if we fully understand the selection effects, which means
it is necessary to model cluster physics as well.  This is in fact where
combining surveys becomes particularly powerful, since we can
measure the density and temperature profiles of clusters (as well as groups,
and even galaxies) through comparing with X-ray and optical data.

Determining the masses of clusters is perhaps the greatest challenge here,
but that can be tackled using detailed studies of individual objects using
everything we can bring to bear.  For example this can be done with
instruments such as MUSTANG \citep{Romero2019}
on the Green Bank Telescope (and similar
instruments on IRAM and the LMT), as well as with the VLA, and eventually
using ALMA in Band~1 (its lowest frequency band; \citealt{Band1}).

\subsection{Kinetic Sunyaev-Zeldovich effect}

The kinetic SZ (or kSZ) effect is much (${\sim}\,10$ times)
weaker than tSZ, and comes from a
combination of optical depth and line-of-sight velocity.  Despite its
inherent weakness, it is potentially extremely powerful in giving us an
unbiased tracer of the peculiar velocity field \citep[see e.g.,][]{Ma_Scott}.
It is effectively tracing the radial peculiar momentum of the gas field and
has been detected through various cross-correlation methods
\citep[e.g.,][]{Hand2012}.  In fact there are several ideas for statistically
extracting kSZ signals that rely on comparison with other data sets, and
these are all essentially different kinds of 3-point functions of maps and
catalogues \citep{KMM}.

Both the kSZ and tSZ effects can be used to trace the (sometimes called
``missing'') baryons that exist in the warm-hot intergalactic medium, which
are hard to detect otherwise \citep[e.g.,][]{vwHM}.
It is also possible to trace
the gas in the filaments of the cosmic web \citep{Tanimura2019}.  Furthermore a
combination of kSZ+tSZ+lensing will enable us to test feedback models
for galaxy formation \citep{Battaglia2019}.

On top of all that, the kSZ effect also
yields new ways to probe the epoch of reionisation, which are very
complementary to the information expected to come from redshifted 21-cm 
studies and searches for the high-$z$ sources of ionization
\citep{patchy}.

\subsection{Cosmic infrared background}

About half of the light emitted by stars is absorbed by dust and reradiated
in the thermal infrared and then redshifted to longer wavelengths
\citep[see e.g.,][]{Hill2018}.
Thus the far-IR/mm regime allows us to track star-formation
(or ${\dot M}_\ast$), to complement the direct stellar light (or $M_\ast$)
that we see in the optical/near-IR.
All of the star-forming galaxies in the Universe
combine to give the cosmic infrared background (CIB).  The contributions to
the CIB can be studied using its statistics, specifically the 1-point
\citep[i.e., histograms][]{BLAST_PofD}
and 2-point \citep[i.e., power spectra][]{Viero2013} functions of the maps.
The CIB power spectrum carries particularly valuable information about the
clustering and redshift distribution of the sources.  With multi-band
measurements of the CIB one can extract tomographic information (and more
can be extracted by cross-correlating with other probes).

The CIB anisotropies have already been measured with much higher S/N than
generally appreciated, using BLAST, {\it Herschel}, {\it Planck}, ACT
and SPT, in many mm-to-submm bands \citep{Viero2019}.
With improved measurements (and models) we will be able to trace how
dark-matter halos turn their baryons into stars {\it and\/} at the same time
constrain background cosmological models.

Peaks in the CIB can be identified as clumps of star-forming galaxies,
which are candidates for ``proto-clusters''.
Huge surveys at these wavelengths enable the discovery of large numbers of
these objects, as has already been carried out with {\it Planck\/}
\cite{PlanckIntXXXIX} and SPT \cite{Miller2018}.  But future surveys
will generate well-understood statistical samples containing thousands of
objects, which can be studied in surveys from other wavebands, and will
be targets for pointed observations with {\it JWST}, TMT, etc.

\subsection{Cross-correlations}

It seems likely that there will be an explosion of activity in comparing
wide-field surveys at disparate wavelengths.  The coming of SKA, LSST,
{\it Euclid}, {\it eROSITA\/} and other huge-scale efforts will make this
inevitable (see white papers by \citealt{SKA_wp} and \citealt{WF_wp}).
There are several different statistical approaches that can be
used here, including ``stacking'' (i.e., the covariance between a map and
a catalogue), cross-correlations between maps and bispectra (between three
data products).  The available data products will include maps and catalogues,
in both 2 and 3 dimensions.
But it is important to point out that exactly what will come out of such
cross-correlation studies is {\it not known\/}!

Combination of tSZ, X-ray and lensing data will yield the density and
temperature profiles around clusters -- statistically for large samples of
objects and in more detail for brighter (or closer) individuals.  This means
we have the ability to look at the effects of feedback (from AGN or SNe,
for example) on the IGM.  This may be the best way to understand the
relationship between galaxy formation and large-scale structure.

21-cm and other intensity-mapping surveys are another area of expertise
for Canada, particularly with CHIME \citep{CHIME}.  Cross-correlations of
intensity-mapping data-cubes with the CMB secondary signatures will be
particularly interesting for understanding how neutral hydrogen is related
to other baryonic components.

\subsection{Further science directions}

Galactic foreground signals will need to be extracted before these cosmological
signals can be investigated.  Then, of course, such signals can be used to
learn more about our Galaxy, including dust, gas and magnetic field
structures.  There are other diagnostics of the Universe
in front of the CMB that can be extracted from CMB surveys.

With huge swaths of sky scanned continuously,
the cadence of such surveys could be nearly daily.  As an example, at a
wavelength of 2\,mm, CMB-S4 could provide 2-mJy rms maps of half the sky
every 2 days.  CCAT-prime could provide similar monitoring at higher frequency.
This would give us monitoring of known sources such as:
GRBs; AGN; blazars; XRBs; novae; protostars; and asteroids.
Additionally, the mm and submm sky could also be searched for other
time-variable phenomena, such as
FRBs, GW sources, SNe, extra planets, and (of course) things we haven't
even thought of yet!
Ambitious CMB-like experiments will thus play a role in multi-messenger
astronomy, bringing the mm waveband into the mix for the first time, which
will be particularly important for obscured sources that could be hard to
detect in the optical or X-ray bands.

With the right frequency channels, it will be possible to extract catalogues of
both virialised clusters (through the tSz effect) and proto-clusters (i.e.\
peaks in the CIB).  Of particular interest will be
understanding how these populations are related, and whether we can find
the intersection, where clumps of star-forming galaxies turn into
actual clusters.

Changes in the CMB anisotropies at high frequencies could indicate the
effects of Rayleigh scattering or resonant line scattering, giving information
on high-redshift modes that could further constrain cosmological parameters.
Beyond the tSZ and kSZ effects, one can also try to use the relativistic
and non-thermal SZ effects to measure gas temperatures and feedback within
clusters, while the polarized SZ effect can measure transverse velocities
\citep[see][for more discussion]{Backlight}.

This white paper has focused on broad-band detectors of continuum radiation.
However, spectroscopic capability at these same wavelengths allows for several
new directions, e.g.\ intensity mapping of CO or {\sc Cii} lines, blind
redshift surveys and tomography of other signatures, as well as studies
of individual extragalactic sources over a wide range of redshifts with
space observatories such as {\it SPICA\/} and {\it Origins\/} \citep[see the
white paper by][]{FIR_wp}.

\section{Conclusions}

The science described here is very well aligned with major interests among
Canadian astronomers.  The example projects given
are generally open to involvement
by any interested Canadian researcher, and indeed there are many who are
already involved.  As a nation of astronomers, we are particularly good at
dealing with wide-field surveys, and the relatively small size of our
community means that we are quite familiar with the benefits of
multi-waveband approaches.  We therefore have some advantages that make
it natural for us to tackle science questions that come from the combination of
signatures from many different components of our Universe.  Those that are
extracted from CMB-type experiments are new and growing in importance.

Despite our advantages as Canadians, the major challenge is competing with
well-funded teams among our collaborators in other countries.  It is
important that there are opportunities for securing sufficient funding for
hardware, software and data analysis contributions to these projects.  This
is the only way that we will be able to play leadership roles in the
scientific exploitation of these enormously exciting data sets.

\clearpage

\bibliography{cmb_non_cmb}

\end{document}